\documentclass{article}

\usepackage{arxiv}
\usepackage[utf8]{inputenc} % allow utf-8 input
\usepackage[T1]{fontenc}    % use 8-bit T1 fonts
\usepackage{hyperref}       % hyperlinks
\usepackage{url}            % simple URL typesetting
\usepackage{booktabs}       % professional-quality tables
\usepackage{amsfonts}       % blackboard math symbols
\usepackage{nicefrac}       % compact symbols for 1/2, etc.
\usepackage{lipsum}
\usepackage{graphicx}
\graphicspath{ {./images/} }

\title{Communication-Induced Bifurcation and Collective Dynamics in Power Packet Networks: A Thermodynamic Approach to Information-Constrained Energy Grids}

\author{
 Takashi HIKIHARA \\
 Kyoto University\\
 Yoshida-Honmachi,  Sakyo, Kyoto 606-8501, JAPAN \\
  \texttt{hikihara.takashi.2n@kyoto-u.ac.jp} \\
 }

\begin{document}
\maketitle
\begin{abstract}
This paper investigates the nonlinear dynamics and phase transitions in power packet network connected with routers, conceptualized as macroscopic \textbf{information-ratchets}. In the emerging paradigm of cyber-physical energy systems, the interplay between stochastic energy fluctuations and the thermodynamic cost of control information defines fundamental operational limits.
We first formulate the dynamics of a single router using a Langevin framework, incorporating an exponential cost function for information acquisition.  Our analysis reveals a discontinuous (first-order) phase transition, where the system adopts a strategic abandon of regulation as noise intensity exceeds a critical threshold $D_c$. This transition represents a fundamental information-barrier inherent to autonomous energy management.  
Here, we extend this model to network configurations, where multiple routers are linked through diffusive coupling, sharing energy between them. We demonstrate that the network topology and coupling strength significantly extend the bifurcation points, with collective resilient behaviors against local fluctuations.  These results provide a rigorous mathematical basis for the design of future complex communication-energy network,  suggesting that the stability of proposed systems is governed by the synergistic balance between physical energy flow and the thermodynamics of information exchange.  It will serve to design future complex communication-energy networks, including internal energy management for autonomous robots.
\end{abstract}


\maketitle

\section{Introduction}
The rapid expansion of renewable energy sources introduces intense stochastic fluctuations into the power grid, manifesting as an influx of environmental entropy. Traditional power systems have maintained stability through margins for physical dissipation, such as mechanical inertia and excessive reserve capacity. 
However, as the penetration of intermittent power sources rises, these conventional stabilization methods are facing their thermodynamic and economic limits.  In this context, the power packet network is one of emerged technologies as a promising paradigm, where energy is discretized into packets associated with information tags (headers) for autonomous routing.

The physics of such a system can be understood through the view of information thermodynamics, 
specifically as a macroscopic implementation of an information-ratchet.
In this framework, power packet routers operate as \textbf{Maxwell’s Demons,}  reducing system entropy by acquiring and consuming information to extract effective work from stochastic fluctuations. 
Previous studies \cite{Toyoda1998, Takuno2010,  Abe2011, Gelenbe2012} have demonstrated the concept of power packet transfer.  The further studies revealed the fundamental physical limit and relationship to information science \cite{Nawata2018,Takahashi2015,Katayama2020, Mochiyama2025}.   The cost of communication and the increasing entropy came out for scaling up the networks.    
Unlike idealized theoretical models, real-world communication and information processing in power packet routers cannot avoid energy dissipation due to computational complexity and high-speed power switching.  

In this paper, we propose the concept of Communication-Induced Bifurcation, a phenomenon where the thermodynamic cost of control information triggers a discontinuous phase transition in the system dynamics. 
We first obtain a model for single router using a Langevin equation with an exponential information processing cost. 
We clarify that when environmental noise exceeds a critical threshold $D_c$, the system undergoes a strategic abandon by control to avoid catastrophic energy dissipation. 
Then, we extend the discussion to networked configurations, for demonstrating how collective dynamics and spatial entropy smoothing through diffusive coupling appear and push these information barriers.  The result enhances the resilience of the grid finally. This theoretical framework provides a new design principle for future possible information-constrained energy grids, where the ultimate stability is governed by the synergistic balance between physical power flow and the thermodynamics of information exchange.  

\section{Mathematical Formulation of Power Packet Networks as Information Ratchets}
\begin{figure}[tbp]
\centering
\begin{tabular}{c}
\begin{minipage}{0.3\textwidth}
\centering
\includegraphics[width=\textwidth]{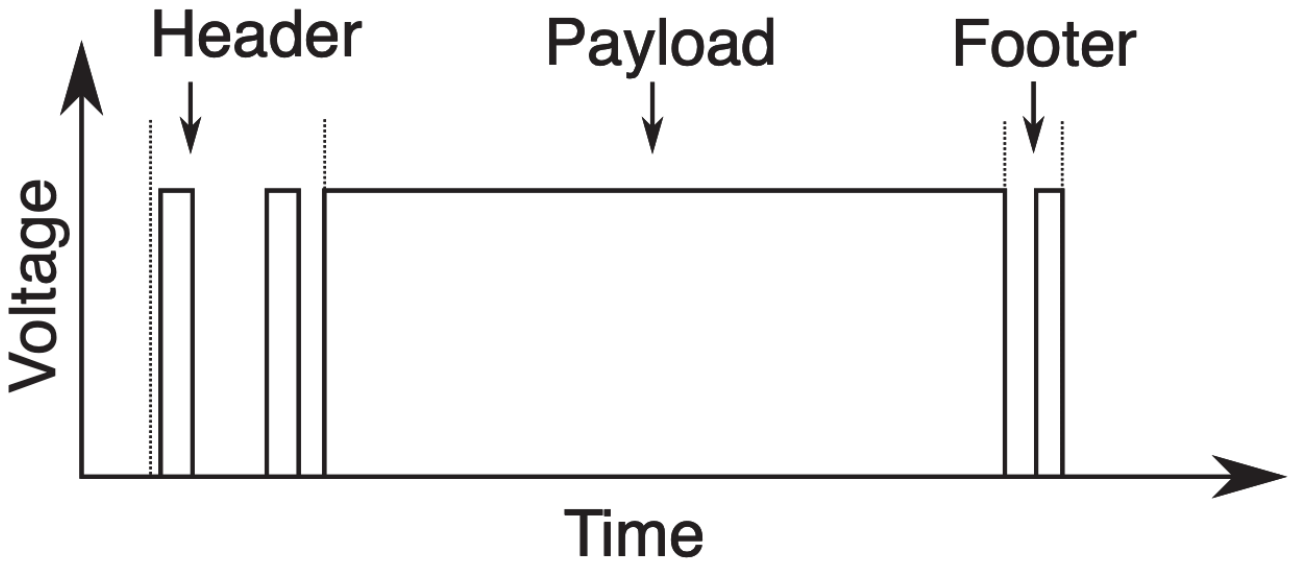}

(a)
 
%\caption{Structure of power packet.}
%\label{fig:packet}
%\end{figure}
\end{minipage}

\begin{minipage}{0.3\textwidth}
%\begin{figure}[tb]
\centering
\includegraphics[width=\linewidth]{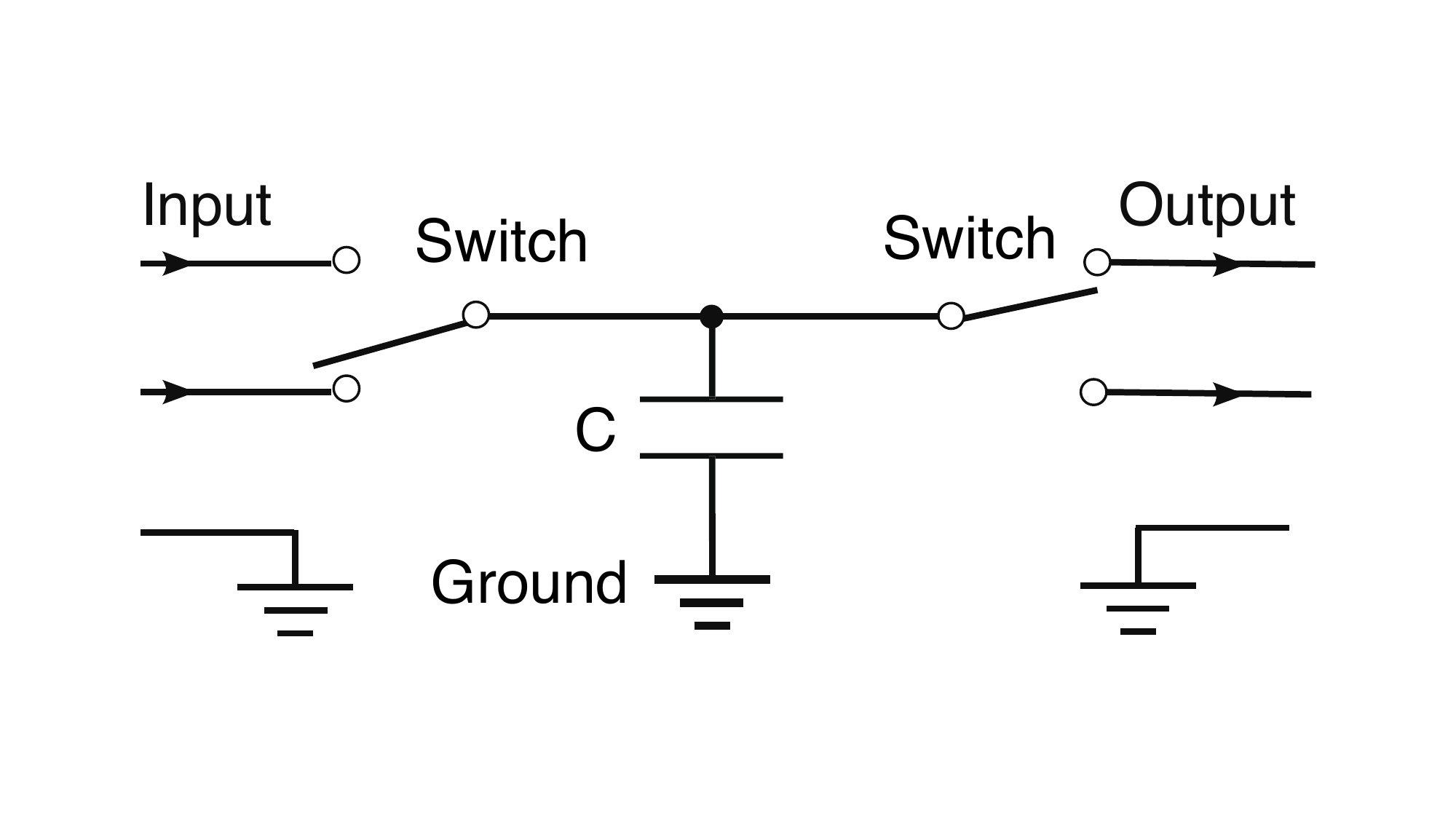}

(b) 

\end{minipage}
\end{tabular}

\begin{minipage}{0.7\linewidth}
\centering
%\begin{figure}[tb]
% \centering
\includegraphics[width=\linewidth]{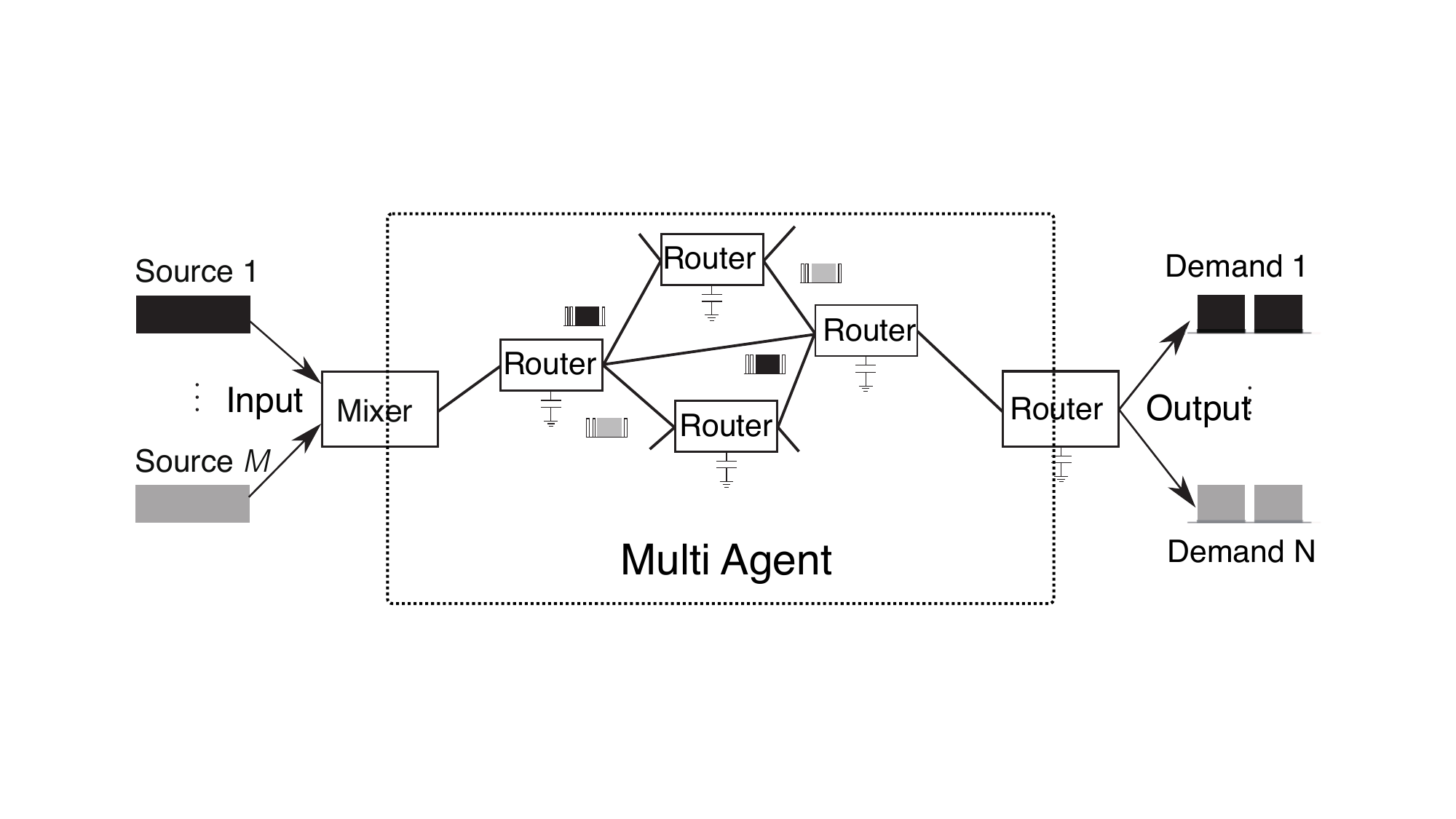}

(c)

\end{minipage}
\caption{Concept of power packet and its physical modeling as an information-ratchet.  
(a) Structure of a power packet, where the information payload (header and footer) is physically integrated and transmitted simultaneously with the energy unit.  
(b) Schematic of a power packet router, which operates as a Maxwell’s demon by processing the integrated information to regulate stochastic energy flows.  
(c) Equivalent dynamical model under information-constrained feedback. The control input $u$ is determined by the information co-transmitted with the packet, incurring a thermodynamic cost $\Phi(u)$ that acts as a nonlinear feedback to the Langevin dynamics.}
\label{fig:router_model}
\end{figure}

We consider the power packet distribution by routers and the network as shown in \ref{fig:router_model}.  The required information for routing co-exists with the energy pulse within the power packet structure (\ref{fig:router_model}(a)).  Consequently, the acquisition of this co-transmitted information inevitably imposes a thermodynamic overhead on the router's dynamics.

\subsection{Physical Model of Power Packet Router}
In a distributed power network using power packets, a node (router) is defined as a non-equilibrium open system with energy state $x \in \mathbb{R}$. Let $H(x_t, \lambda_t)$ be the Hamiltonian of the system at time $t$. Here, $\lambda_t \in \{0, 1\}$ is a control parameter representing the packet selecting operation (open/close of a switch) in the router.  The dynamics of the input power can be expressed by the following Langevin equation, including stochastic fluctuations from the supply side:
\begin{equation}
\frac{dx_t}{dt} = -\nabla H(x_t, \lambda_t) + \sqrt{2D} \xi(t)
\end{equation}
Here, $x$ denotes the energy storage level (buffer energy) within the router, and $\dot{x}$ governs the dynamics of the active power flow.  $-\nabla H(x_t, \lambda_t) $ denotes the energy dissipation in router.  $\xi(t)$ is white noise with mean 0 and variance 1, and $D$ represents the noise intensity (diffusion coefficient) from the environment, which reflects the intermittency of renewable energy supply depending on weather conditions.  In the following discussion, $u$ is the optimized rate of selection of $\lambda_i \in \{0,1\}$.  The hardware realization of this switching, synchronized with the information tag, was established for bidirectional flows in \cite{Mochiyama2025}.  

%%%%%%%%%%%%%%%Fig1%%%%%%%%%%%%%%%
%Fig.1 
%%%%%%%%%%%%%%%%%%%%%%%%%%%%%%

\subsection{Information Ratchet Mechanism and Mutual Information}
In power packet networks \cite{Takahashi2015, Sugiyama2015, Katayama2020, Nawata2018, Mochiyama2025}, the operation of a router can be understood as an information ratchet, an engineering implementation of Maxwell's Demon. The router feedback-controls the switch $\lambda_t$ at appropriate timings based on observations of supply side fluctuations to reduce the system's entropy and provide a stable power flow required by the demand side. In this process, the following inequality, an extension of the Sagawa-Ueda relation, holds between the mutual information $I$ obtained from observation and the effective work $W$ output by the system \cite{Sagawa2008}:
\begin{equation}
\langle W \rangle \le -\Delta F + kT \langle I \rangle
\end{equation}
where $\Delta F$ is the change in free energy. In the power packet network of this study, $kT \langle I \rangle$ on the right side provides the theoretical upper limit for the improvement of power quality extracted by information.

\subsection{Exponential Communication and Information Processing Cost Model}
In real routers, the acquisition and erasure of information are accompanied by unavoidable energy dissipation. 
Reflecting the physical complexity and high-speed switching losses observed in physical routers equipped with SiC MOSFETs and FPGA control units [5, 8], we model the dissipation cost $\Phi(u,D)$ for maintaining control effort $u\in[0,1]$ as:
\begin{equation}
\Phi(u, D) = \kappa \cdot D \cdot \left( \exp(\beta u) - 1 \right)
\end{equation}
In this equation, the coefficient $\kappa$ represents the computational complexity required to maintain the system's demand response, and $\beta$ represents the dissipation constant per unit noise.

The choice of the exponential form is physically justified by the nonlinear growth of switching frequency required to suppress high-level fluctuations and the algorithmic complexity of real-time estimation under intense noise. If a polynomial or logarithmic function were employed instead, the information gain would unphysically dominate the cost even in an infinitely rough environment, violating the generalized second law of thermodynamics. Therefore, the characteristic that the cost increases exponentially relative to the product of $D$ and $u$ becomes a critical physical factor that limits information processing in high-noise environments.

\subsection{Definition of System Evaluation Function}
To integrally evaluate the {\bf quantity} of energy, its entropic {\bf quality}, and the {\bf information cost}, we introduce the following evaluation function $J$:
\begin{equation}
J(u) = \alpha \cdot G(u) - \Phi(u, D) - T\Delta S
\end{equation}
Here, $G(u) = 1 - \exp(-\gamma u)$ is the gain function for satisfying demand through control, $\alpha$ is the market or physical value of energy quality, and $T\Delta S$ is the loss penalty due to residual entropy. Searching for the $u^*$ that maximizes $J(u)$ reduces to a thermodynamic optimization problem in power packet networks.  
Note that all terms in the objective function $J(u)$, including the control gain and the information cost $\Phi$, are defined in terms of energy rates (Watts, $J/s$) to ensure dimensional consistency with the power flow dynamics $\dot{x}$.  Specifically, the dimensional exchange  between the energy per packet bound in Eq. (2) and the continuous rate objective $J(u)$ is maintained by multiplying the thermodynamic terms by the average packet frequency $\nu$, converting Joules into Watts.  

\section{Numerical Calculation Algorithm and Optimization Process}
In this report, we sequentially determine the control amount $u^*(t)$ that maximizes the net evaluation value $J$ of the system under dynamically changing environmental noise $D(t)$. To maximize $J(u)$ under a non-stationary environment $D(t)$, the following steps are executed sequentially:

\subsection{Sequential Estimation Algorithm for Environmental Noise}
The router observes input power fluctuations in real time and estimates the local noise intensity (diffusion coefficient) $D(t)$. At discrete time steps $\Delta t$, the following moving average is calculated based on the differences in the observed energy state $x$:
\begin{equation}
\hat{D}(t) = \frac{1}{2 \Delta t} \left\langle [x(\kappa) - x(\kappa-1)]^2 \right\rangle_{W}
\end{equation}
where $\langle \cdot \rangle_{W}$ represents the sample average over a time window of length $W$. This estimated value $\hat{D}$ serves as a parameter for the optimization in the next section.

\subsection{Maximization of Thermodynamic Evaluation Function}
At each sampling time, the following nonlinear optimization problem is solved:
\begin{equation}
u^*(t) = \mathop{\rm arg~max}\limits_{u \in [0, 1]} \left\{ \alpha G(u) - \kappa \hat{D}(t) (e^{\beta u} - 1) - T \Delta S(u) \right\}
\end{equation}

\subsection{Physical Mapping of Control Parameters}
The optimized $u^*(t)$ is converted into the switching operation of a physical power packet router. Packet generation is considered to be the discretization of input energy into constant quanta $\Delta E$. The duty ratio of the switch can be converted from the calculated $u^*$ into the operating rate (effective packet selection rate) of the unidirectional switch corresponding to the ratchet pawl.  Dissipation corresponds to subtracting the actually consumed computational energy $\Phi(u^*, \hat{D})$ from the system's state variables.

\section{Numerical Experimental Results and Physical Considerations}
\subsection{Setting the Randomness of Pseudo-Solar Output}
\begin{figure}[b!]
\centering
\includegraphics[width=3.3in]{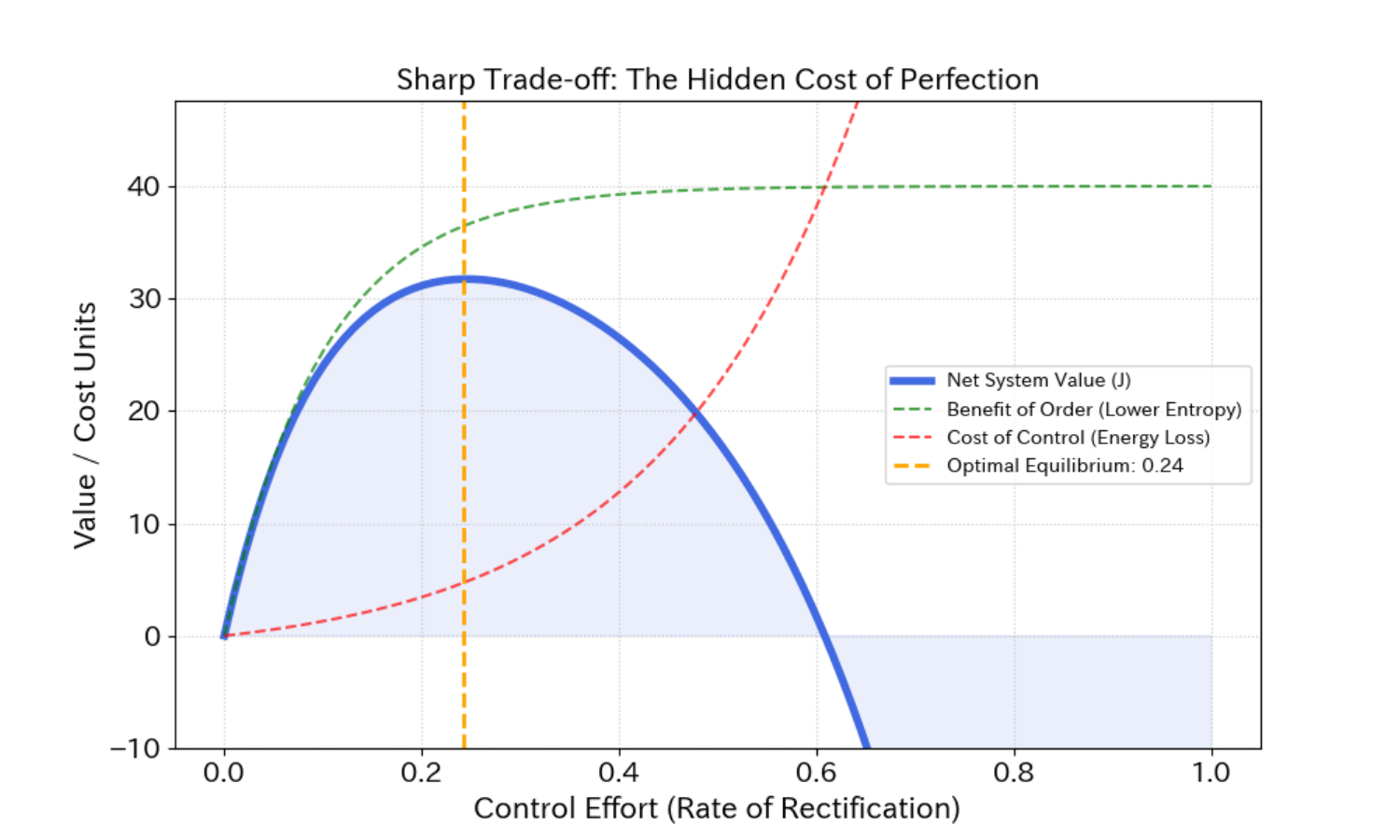}
\caption{Thermodynamic trade-off between energy and information. Profile of the system evaluation function $J$ (thick blue line) against control amount $u$. While the gain for satisfying demand response (dotted green line) saturates, the information processing cost (dotted red line) increases exponentially. It is shown that the global optimal solution $u^*$ (dotted yellow line) is obtained in a region where energy and entropy are balanced.}
\label{fig:trade_off}
\end{figure}

In this section, we verify the effectiveness of the proposed optimization algorithm. Input power is given stochastic intermittency by a stochastic process modeling solar power generation. White noise with variance $D(t)$ is superimposed on a steady expected output, and this $D(t)$ is used as a control parameter defining the environmental randomness (entropy influx intensity) to scan the response. As shown in Fig. \ref{fig:trade_off}, increasing the control amount $u$ causes the gain for meeting demand response to saturate, while the information processing cost increases exponentially. From these asymmetric physical characteristics, it is clear that an optimal solution $u^*$ exists for the system that balances energy reserves with entropic quality.

%%%%%%%%%%%%%%%Fig2%%%%%%%%%%%%%%%

\subsection{Autonomous Adaptation to Environmental Noise}
\begin{figure}[t!]
\centering
\includegraphics[width=3in]{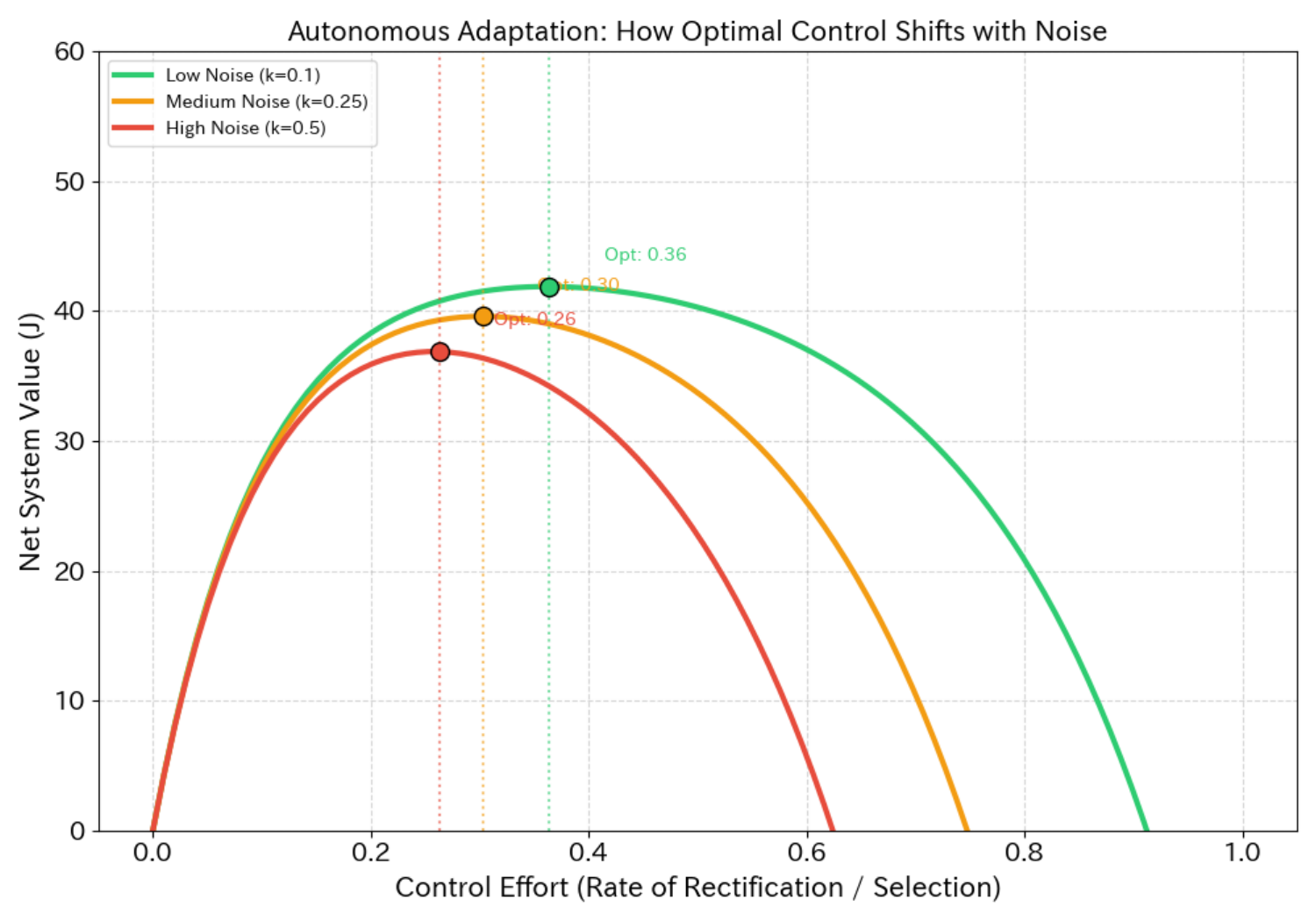}
\caption{Autonomous adaptation of control to environmental noise. Showing the change in evaluation function $J$ with increasing environmental noise intensity $D(t)$. As noise becomes more severe, dissipation costs associated with information processing become dominant, and the system autonomously shifts $u^*$ to the low-output side to avoid excessive energy consumption.}
\label{fig:adaptation}
\end{figure}

Fig. \ref{fig:adaptation} shows the change in the profile of the system evaluation function $J$ as the environmental noise intensity $D(t)$ changes. As noise intensity increases, the peak (optimal point) of the evaluation function curve shifts continuously to the lower side (left). This quantitatively represents the behavior where, as the environment becomes rougher, the cost of information to satisfy demand jumps, and the system autonomously relaxes control to suppress energy consumption.

%%%%%%%%%%%%%%%Fig3%%%%%%%%%%%%%%%

\subsection{Analysis of Discontinuous Phase Transition at Critical Noise $D_c$}
\begin{figure}[b!]
\centering
\includegraphics[width=0.4\textwidth]{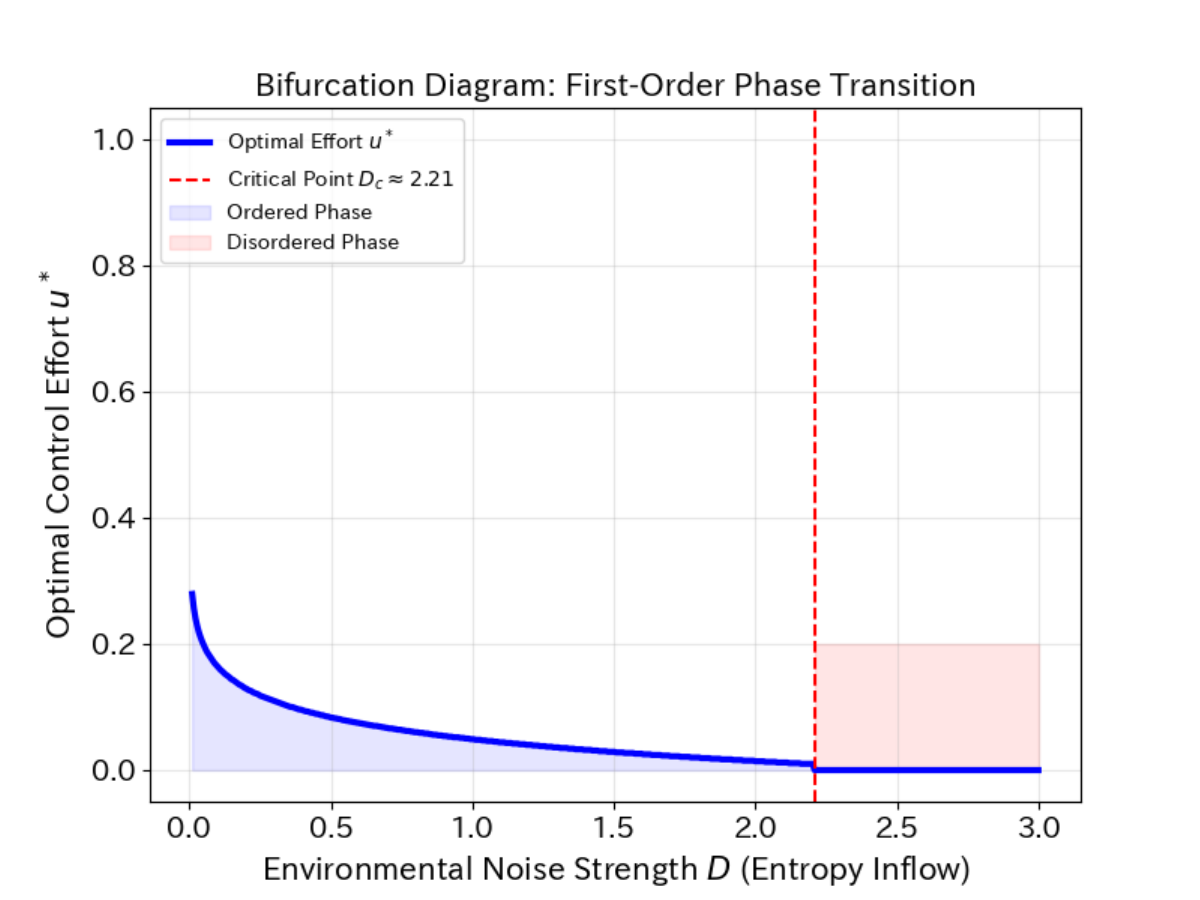}
\caption{Bifurcation diagram of discontinuous phase transition. Showing a discontinuous (first-order) phase transition where $u^*$ drops vertically to 0 the moment noise $D(t)$ exceeds the critical value $D_c$. At the critical value $D_c \approx 2.21$, the system shifts discontinuously from an ordered phase to a disordered phase. This corresponds to a state where the system autonomously abandons control to provide thermodynamic suppression because the dissipation cost of information has overwhelmed the gain of order formation.}
\label{fig:figure_bifurcation}
\end{figure}

This section analyzes the communication-induced bifurcation, where the thermodynamic cost of information exchange triggers a discontinuous transition.  
As environmental noise $D(t)$ increases, a discontinuous transition occurs in the control response. As the bifurcation diagram in Fig. \ref{fig:figure_bifurcation} shows, the moment the noise intensity reaches a critical value $D_c \approx 2.21$, the optimal control effort $u^*$ jumps discontinuously from a finite value (approx. 0.01) to 0. In the ordered phase ($D < D_c$) before reaching the critical point, the system functions as Maxwell’s Demon, actively continuing to flow packets in one direction. This is because the gain obtained for satisfying demand exceeds the exponential dissipation cost associated with information processing. However, when noise $D(t)$ exceeds the critical value $D_c$, the maximum of the evaluation function $J(u)$ disappears, or the evaluation value of information falls below the evaluation function at the origin, resulting in thermodynamic failure. At this time, the system enters an autonomous suppression operation, transitioning to a state that preserves stored energy to avoid self-destruction due to excessive dissipation.  The dissipation corresponds to energy loss for information rewriting and erasure and power switching.  

The phase transition stems from a nonlinear trade-off between the cost of acquiring information and the gains of establishing order.  
Although the discontinuity is mathematically driven by the ${\rm arg~max}$ optimization of $J(u)$, it fundamentally represents a first-order phase transition in non-equilibrium thermodynamics. The evaluation function $J(u)$ acts as a thermodynamic potential where the order parameter $u^*$ experiences a sudden collapse as the global maximum switches abruptly from the ordered phase to the disordered phase at the critical parameter $D_c$.  

%%%%Fig4
%20260317

%%%%%%%%%%%%%%%Fig5%%%%%%%%%%%%%%%
\begin{figure}[htbp]
\centering
    %\rule{0.8\linewidth}{4cm} %final_solar_loss_optimization.png
\includegraphics[width=3.1in]{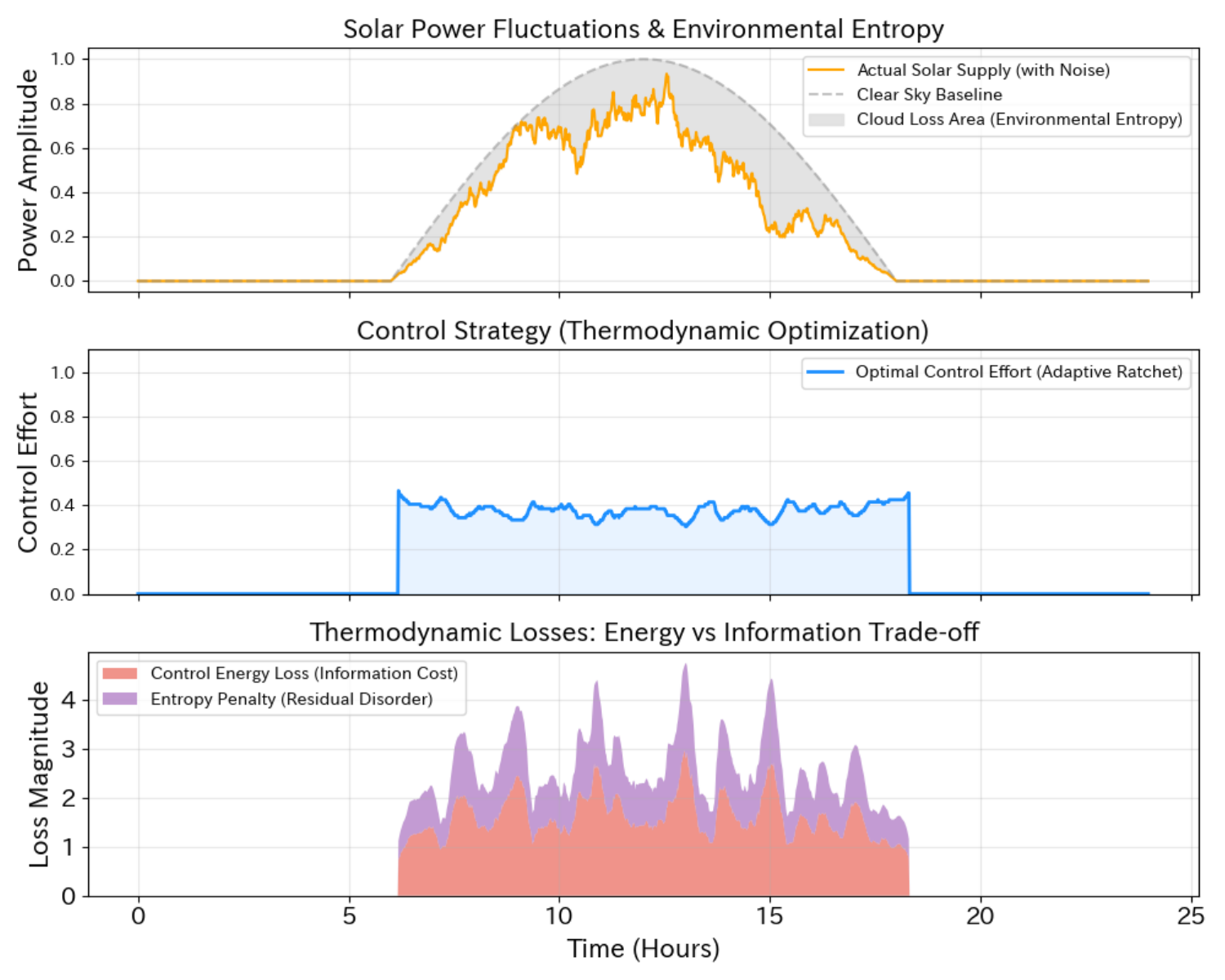}
\caption{Dynamic optimization and loss decomposition under pseudo-solar supply by adaptive control strategy under communication constraints.   (Top) Fluctuations in pseudo-solar output and the resulting transition of environmental entropy influx intensity. (Middle) Real-time transition of optimal control effort $u^*$ (adaptive ratchet) according to estimated noise intensity. (Bottom) Stacked chart showing the dynamic balance between energy dissipation due to information processing (red) and quality loss due to residual entropy (purple).}
\label{fig:final_solar_loss_optimization}
\end{figure}

\subsection{Thermodynamic Superiority of the Proposed Control Method}
The proposed adaptive control is applied to the dynamic optimization and loss decomposition under pseudo-solar supply in Fig. \ref{fig:final_solar_loss_optimization}, which corresponds to the dynamic control reduction in the adaptive strategy (Fig. \ref{fig:final_solar_loss_optimization} (Middle)) .  The drop in control amount observed at the noise peak physically demonstrates the emergence of a discontinuous phase transition (first-order phase transition) due to thermodynamic adaptation in harsh environments.

In conventional control, information costs increase exponentially under high noise, causing the system's evaluation function $J$ to fall into a negative region, which is a region where control becomes counterproductive. In contrast, the proposed adaptive ratchet treats the critical point $D_c$ as a physical limit, dynamically adjusting information consumption and reducing the burden in harsh environments to maximize operational feasibility.  

\section{Cooperative Behavior in Multi-Agent Systems}
\subsection{Network Model with Diffusion Coupling}
Having shown the bifurcation behavior of a single router, this section extends the system to multiple routers exchanging packets between adjacent nodes. For the energy balance of node $i$, a diffusion coupling term using a coupling constant $g$ with the set of adjacent nodes $N_i$ is introduced:
\begin{equation}
\frac{dx_i}{dt} = f(x_i, \lambda_i) + g \sum_{j \in N_i} (x_j - x_i) + \sqrt{2D_i} \xi_i(t)
\end{equation}
where $f(x_i, \lambda_i)$ represents the deterministic contribution based on the previously mentioned Hamiltonian. 
%In the following discussion, $u$ is the optimized rate of selection of $\lambda_i \in \{0,1\}$.  
%
The coupling constant $g$ represents the communication-induced energy sharing, which depends on $x_j -x_i$.  It spatially dissipates energy and entropy from high-noise nodes to low-noise nodes.

\subsection{Extension of Critical Points as a Collective Phenomenon}
The discontinuous phase transition identified in the single node represents the thermodynamic limit when operating routers independently. However, in the multi-agent network proposed in this paper, energy and entropy are shared between adjacent nodes, making it possible to dynamically extend this critical point $D_c$. An entropy smoothing effect arises due to the network topology; nodes exposed to locally intense noise distribute the effective load to adjacent low-noise nodes via diffusion coupling $g$. This interaction spatially relaxes the information barriers faced by individual nodes, pushing the single-node critical point $D_{c,single}$ to a higher noise intensity $D_{c,network}$ and triggering collective phenomena.

\begin{figure}[htbp]
\centering
\includegraphics[width=3.3in]{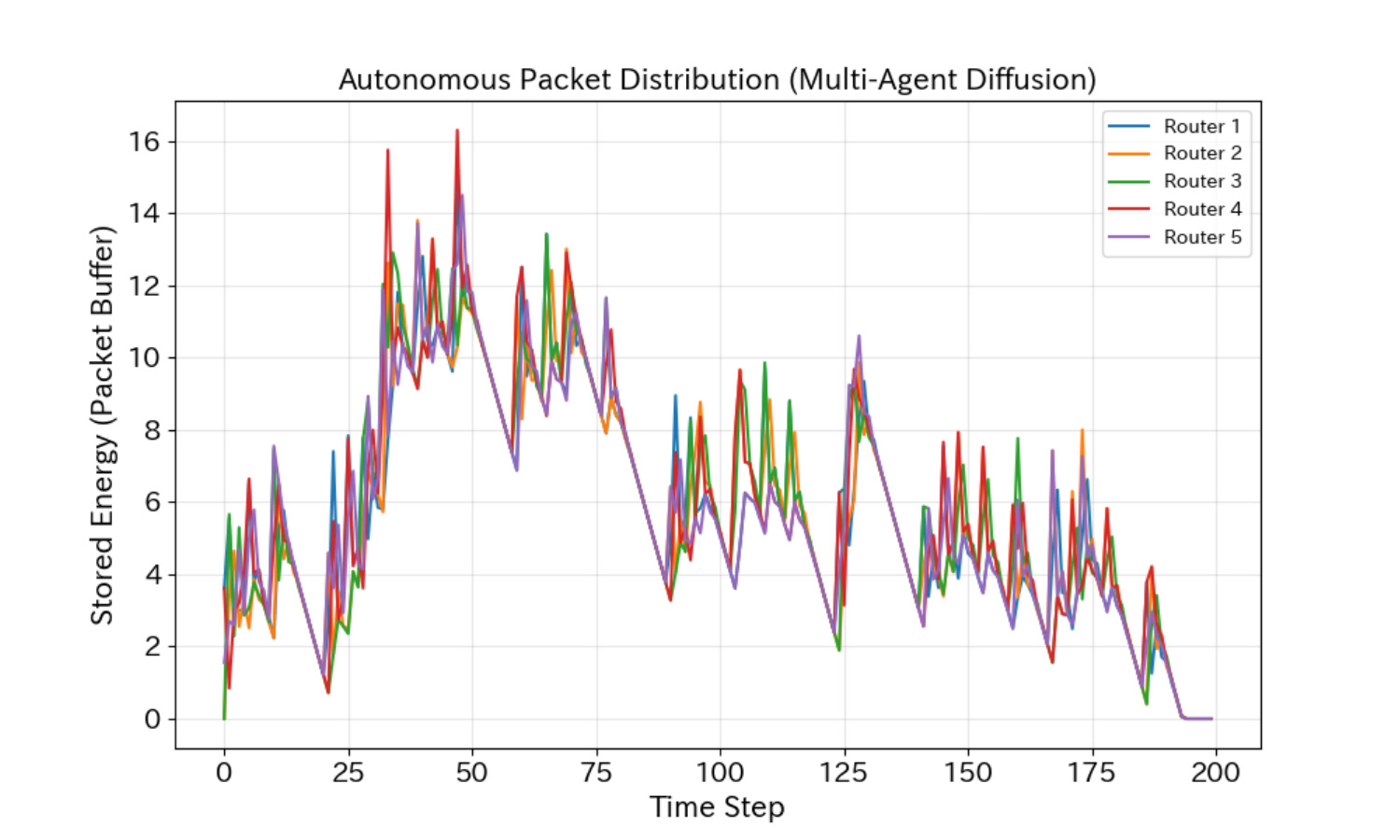}
\caption{Autonomous decentralized power packet exchange and spatial smoothing of entropy in multi-agent systems. It sures a collective resilience emerging from networked communication.  In a networked five routers with diffusion coupling $g$, the transition of stored energy in buffer for each node is obtained. Sudden supply fluctuations (entropy influx) occurring at specific nodes are distributed and shared across the entire network through autonomous packet exchange between adjacent nodes. This collective phenomenon suppresses the risk of individual nodes reaching the information critical point and abandoning control, which implies phase transition, thereby expanding the operating range.}
\label{fig:network_effect}
\end{figure}

Fig. \ref{fig:network_effect} shows the simulation results for the transition of stored energy in the routers. It is observed that collective order maintenance is clearly emerging. Even in high-noise regions where individual nodes would be forced to suppress control, they operate cooperatively through coupling, maintaining a low-entropy state for the system as a whole. This clarifies that the power packet network goes beyond a simple physical transmission infrastructure to share information dissipation costs and collectively maintain order, emerging as a collective function in terms of information thermodynamics.
%20260317

%%%%%%%%%%%%%%%Fig6%%%%%%%%%%%%%%%

\section{Discussion}
Furthermore, the relationship between the noise intensity $D$ and the control input $u$ suggests an important extension of the fluctuation-dissipation theorem (FDT) within this information-ratchet framework. 
In conventional physical systems, FDT reveals the spontaneous fluctuations of a system to its linear response through temperature $T$.  In our model, the environmental noise $D$ behaves as an effective temperature.  We introduced the information processing cost $\Phi(u, D)$ which represents a computational dissipation and in not considered in classical passive systems. 
The emergence of the critical point $D_c$ can be interpreted as a breakdown point of the steady-state balance between energy extraction and this information-induced dissipation. 
This formulation provides a perspective on how the generalized second law of thermodynamics governs the operational limits of cyber-physical systems.  It will be the way for a unified variational principle for networked information-thermodynamic system.

\section{Conclusion}
This report redefines the physical essence of energy transport in distributed power networks using power packets as an information-ratchet mechanism from the perspective of non-equilibrium statistical mechanics. We formulated the process where the stochastic fluctuations of the supply side, accompanying the expansion of renewable energy, are treated as environmental entropy influx, which power packet routers control through the acquisition and consumption of information \cite{Inagaki2021}.

Through numerical experiments, it was clarified that an unavoidable thermodynamic trade-off exists between the quantity of energy and its entropic quality via information processing costs. As Fig. \ref{fig:trade_off} shows, while the gain for satisfying orders saturates, the dissipation cost associated with information processing increases exponentially, which can act as a physical barrier. This reveals that fixed stabilization does not necessarily maximize the system's evaluation function and that information-thermodynamic constraints arise.

Furthermore, it was clarified that a discontinuous phase transition, first-order phase transition, occurs when environmental noise intensity exceeds a critical value $D_c$. To prevent the free energy from falling into the negative under harsh fluctuations, the system chooses to autonomously abandon control and minimize dissipation; thus, the dynamic reduction in control amount observed at noise peaks in Fig. \ref{fig:final_solar_loss_optimization} can be a thermodynamically honest adaptation. Finally, confirmed effects of diffusion coupling in multi-agent systems showed that spatial smoothing of entropy by network topology pushes up individual critical points and generates collective phenomena for the whole system.

%We clarify that $J(u)$ serves as a 
We suggest that $J(u)$ can be interpreted as a prospective 
generalized potential that incorporates the rate of free energy change and information flow within the framework of non-equilibrium thermodynamics. Then, we will add a discussion regarding the prospective generalization of this theory, specifically its reduction to a variational principle. 
It indicates a potential paradigm shift in the design of power networks from conventional stabilization relying on physical margins to thermodynamic optimization based on information. In power packet networks that guide energy by consuming information, power packet routers demonstrate the behavior of Maxwell's Demon through the process of entropy reduction.  
This framework is particularly relevant for the internal energy management of autonomous robots, in which communication bandwidth and energy reserves are strictly constrained.

%%%%%%%
%\dataavailability
%The simulation codes which generate the data related to this paper are available from 
%\url{https://doi.org/10.57723/kds618534}. 
%
%\funding
%This work was financially supported  in part by KAKENHI（Grants-in-Aid for Scientific Research (C)) 26K07513.  
%
%\conflictsofinterest
%The author declares no competing interests.
%
%\authorcontribution
%The author TH contributes all process of writing this paper, including conceptualization, investigation, simulations, visualization, and funding.  
%
%\aitools
%Author uses AI tool (Gemini, Gemma) to fix the coding errors and checking logic of paper.  
%
%%%%%%%

\end{document}